\documentclass[twocolumn,aps,prd,amsmath,amssymb,showpacs,preprintnumbers]{revtex4-1}
\usepackage{graphicx,hyperref,color}

\begin{document}

\newcommand{\bea}{\begin{eqnarray}}
\newcommand{\eea}{\end{eqnarray}}
\renewcommand{\d}{{\mathrm{d}}}
\renewcommand{\[}{\left[}
\renewcommand{\]}{\right]}
\renewcommand{\(}{\left(}
\renewcommand{\)}{\right)}
\newcommand{\nn}{\nonumber}

\newcommand{\be}{\begin{equation}}
\newcommand{\ee}{\end{equation}}
\newcommand{\Su}{\color{red}}

\title{The no-boundary wave function for loop quantum cosmology}

\author{Suddhasattwa Brahma}
\email{suddhasattwa.brahma@gmail.com}
\affiliation{The Asia Pacific Center for Theoretical Physics, Pohang 37673, Korea}

\author{Dong-han Yeom}
\email{innocent.yeom@gmail.com}
\affiliation{The Asia Pacific Center for Theoretical Physics, Pohang 37673, Korea}
\affiliation{Department of Physics, POSTECH, Pohang 37673, Korea}

\begin{abstract}
Proposing smooth initial conditions is one of the most important tasks in quantum cosmology. On the other hand, the low-energy effective action, appearing in the semiclassical path integral, can get nontrivial quantum corrections near classical singularities due to specific quantum gravity proposals. In this article, we combine the well-known no-boundary proposal for the wavefunction of the universe with quantum modifications coming from loop quantum cosmology (LQC). Remarkably, we find that the restriction of a `slow-roll' type potential in the original Hartle-Hawking proposal is considerably relaxed due to quantum geometry regularizations. Interestingly, the same effects responsible for singularity-resolution in LQC also end up expanding the allowed space of smooth initial conditions leading to an inflationary universe.
\end{abstract}

\maketitle

\textit{Motivation.--} Any theory of quantum cosmology must provide dynamical equations for the wavefunction of the universe as well as suitable boundary conditions for it. A well-known example for such a boundary condition is due to Hartle and Hawking (HH) \cite{Hartle:1983ai}
\begin{eqnarray}
	\Psi^{\mathrm{HH}} [h_{ab}, \chi] := \int^{(h,\chi)} \mathcal{D}[g]\,\mathcal{D}[\varphi]  \, e^{-S_{\mathrm{E}}[g,\varphi]/\hbar}\,.
\end{eqnarray}
HH defines an initial state for the universe such that the path integral is over Euclidean $4$-metrics bounded only by the $3$-metric, $h$, and the value of matter field on this hypersurface,  $\chi$. Traditionally, the weighting factor is given by (Euclidean) Einstein gravity, coupled to matter fields $\varphi$.

Loop quantum cosmology (LQC), on the other hand, is the application of quantization techniques from loop quantum gravity (LQG) to minisuperspace cosmological settings \cite{Bojowald:2006da}. In this setting, the low-energy effective action is complemented by \textit{quantum geometrical corrections} \cite{Ashtekar:2003hd}. Canonically, it replaces the usual Wheeler de-Witt differential equation by a `difference equation', of finite step-size, due to a \textit{regularized} Hamiltonian constraint. Obviously, difference equations also require boundary conditions to extract specific solutions. Thus, the question of a suitable initial state remains important even for LQC \cite{Dhandhukiya:2016rrp}\footnote{Claims pertaining to LQC providing a `dynamical' initial condition \cite{Bojowald:2001xa} remain confined to simplest isotropic and homogeneous minisuperspace models, requiring extra input in the presence of cosmological perturbations, or even anisotropies.}. 

In simple models of LQC, one usually chooses a semi-classical (nearly-Gaussian) state at late times, peaked about some classical trajectory, and evolves it backwards to derive a bouncing solution \cite{Ashtekar:2011ni}. This does not, however, imply a \textit{deterministic} bounce, as is sometimes incorrectly \textit{assumed}, since a large portion of the state space remains unexplored. While conclusions regarding singularity-resolution due to bounded curvatures and energy densities can be somewhat generically established in LQC \cite{Bojowald:2006da}, the precise mechanism, naturally, remains intricately tied to the initial state one chooses. In this article, we show that the original HH proposal, adapted to LQC, can lead to exciting new possibilities for a non-singular quantum completion of inflation, profoundly expanding the range of allowed initial values for the scalar field.

Our starting point will be the path integral formulation of LQC \cite{Ashtekar:2009dn,Ashtekar:2010gz,Huang:2011es}, with paths weighted not by the Einstein-Hilbert (EH) action, but a different one due to quantum geometry effects \footnote{There are scenarios (beside LQC) in which the path integral receives (non-perturbative) quantum corrections to the classical action, such as for a particle on a curved manifold \cite{DeWitt:1957at}.}. However, in the (relatively) low-energy limit, below $\mathcal{O}(10^{-3})$ Planck density, this action reduces to the standard EH one very rapidly \cite{Ashtekar:2011ni}. Requiring a no-boundary like wavefunction results in two main findings for this scenario. Firstly, this provides a (topological) principle of setting initial conditions in LQC which would become important in avoiding \textit{ad hoc} choices while dealing with, say, cosmological perturbations. More importantly, the LQC `effective' action, appearing in the no-boundary wavefunction, leads to finite probabilities for Lorentzian histories corresponding to an extended parameter space for inflation, due to novel instantonic solutions which were non-existent in the EH case. 

\textit{Formalism.--} In this article, we shall follow the original HH proposal (with the prescribed contour \cite{Halliwell:1988ik,Halliwell:1989dy}), albeit corresponding to the LQC `effective' action and not the EH one, explicitly demonstrating the existence of new saddle-points which are different from the HH solution. Recently, it has been shown  \cite{Feldbrugge:2017kzv} that the Lorentzian path integral in EH quantum cosmology, analyzed using Picard-Lefshetz theory, does not have the any contribution from the Euclidean saddle points proposed by HH (for a different perspective, see \cite{DiazDorronsoro:2017hti}). Although, in this work, we shall not work with the Lorentzian path integral, this can easily be done for LQC and shall be pursued in future work. 

On the LQC side, a technical gap we fill is to include non-perturbative expressions for `inverse-triad' modifications in the effective action appearing in the path integral. Previously, `holonomy modifications' were taken into account since they are primarily responsible for weighting the paths in a way so as to achieve singularity resolution \cite{Ashtekar:2010gz}. However, we shall show that the inverse-triad corrections are essential for having well-defined (Euclidean) instantons, for a no-boundary like wavefunction, near $a\rightarrow 0$. Rather remarkably, we find that due to these corrections, there exists solutions in scenarios where there were no well-defined instantons in the EH case, beyond the pure de-Sitter (dS) case. 

The system under investigation is the closed FLRW cosmology $\d s^2 = -N^2(t) \d t^2 + a^2 \d\Omega^2$, with $\Omega$ being the metric on a unit 3-sphere. For the $k=1$ case, inverse-triad corrections are particularly relevant and provide significant modifications to LQC dynamics \cite{Corichi:2013usa}. For the matter contribution (generically denoted by $\varphi$), we shall first choose to have only a positive cosmological constant $\Lambda$, and later a free scalar field $\phi$. From now on, we shall use the convention that $\hbar = c = G = 1$.

Our no-boundary proposal for LQC (LQCNB) is given by
\begin{eqnarray}\label{LQCNB}
	\Psi^{\text{LQC}}_{\text{NB}} =\int \d N\int \mathcal{D}a\,\mathcal{D}\varphi\, e^{-S^{\text{LQC}}_{\text{E}}[a(\tau), \varphi(\tau)] }\,,
\end{eqnarray}
with the (Euclidean) LQC effective action, in its phase space version ($\mathcal{H}^{\text{LQC}}$ being the LQC Hamiltonian constraint), written as 
\begin{eqnarray}
	S^{\text{LQC}}_{\mathrm{E}} = \int_0^1 \d \eta L_{\mathrm{E}} = \int \d \eta \left(p_{a}\dot{a} - N \mathcal{H}^{\text{LQC}}_{\mathrm{E}} \right)\,.
\end{eqnarray}
One typically chooses the Euclidean time parameter to run from $0$ to $1$. However, following \cite{Hartle:2008ng}, we can introduce a parameter $\tau(\eta) := \int_0^\eta \d\eta' N(\eta')$, such that one takes the integral over (complex) $\tau$ to go from $\tau_i$ to some $\tau_f$. This is equivalent to choosing the proper time gauge by setting $N=1$.

Since $\mathcal{H}^{\text{LQC}}_{\mathrm{E}}$ is zero on-shell, and the canonically conjugate momentum of the scale factor $a$ is given by $p_a = -3\pi a\dot{a}/2$ \footnote{By an abuse of notation, from now on we shall use the dot `$.$' to refer to $\d/\d\tau$ instead of $\d/\d\eta$. }, we can rewrite the action (on-shell) as
\begin{eqnarray}\label{EffectiveAction}
	S^\text{LQC}_{\mathrm{E}} = - \frac{3\pi}{2} \int a \dot{a}^{2} \d\tau = - \frac{3\pi}{2} \int_{0}^{\tilde{a}} a \sqrt{|\mathcal{V}(a)|} \d a,
\end{eqnarray}
where $\dot{a}^{2} = - \mathcal{V}(a)$. Note that this type of an equation, denoting a real Euclidean instanton as opposed to `fuzzy' complex instantons \cite{Hartle:2008ng,Hwang:2012mf,Hwang:2012zj}, can be obtained only for the specific choices of matter we make in this article -- pure dS and a massless scalar field. In this form, the LQCNB can be written in the steepest descent approximation as 
\begin{eqnarray}
\Psi^{\text{LQC}}_{\text{NB}}  \sim e^{-S^{\text{LQC}}_{\text{E}}[\tilde{a}, \tilde{\varphi}]},
\end{eqnarray}
where $\tilde{a} = a(\tau_f)$\footnote{What we call $\tilde{a}$ in our analysis is the boundary at which $\mathcal{V}$ turns negative, and therefore it coincides with the minimum scale factor $a_\text{min}$ of the `bounce' picture in LQC.}. By considering the effective Friedmann equation for such a model, one can calculate the $\mathcal{V}(a)$ required to completely evaluate this state.

\textit{Pure de Sitter.--} Following \cite{Ashtekar:2006es}, we can write the effective Friedmann equation (in $N=1$ gauge) for $k=1$ LQC as 
\begin{eqnarray}\label{EffFriedmann}
-\dot{a}^{2}=\mathcal{V} := \frac{8\pi a^{2}}{3} f^2(a) \left[\frac{\rho}{f(a)} - \rho_1\right] \left[\frac{\rho_2 - \frac{\rho}{f(a)}}{\rho_c} \right],
\end{eqnarray}
where
\begin{eqnarray}
f(a) &:=& \frac{1}{2} \huge\big| |v(a) - 1| - |v(a)+1| \big|\, ,\\
\rho_1 &:=& -\rho_c \left[\sin^2(\sqrt{\Delta}/a) - (1+\gamma^2) \frac{\Delta}{a^2} \right]\, ,\\
\rho_2 &:=& \rho_c \left[\cos^2(\sqrt{\Delta}/a) + (1+\gamma^2) \frac{\Delta}{a^2} \right]\, ,\\
\rho &:=& \frac{\Lambda}{8\pi}\,,
\end{eqnarray}
with the LQC critical density $\rho_c = \frac{3}{8\pi \gamma^2 \Delta}$ and area-gap $\Delta = 2\sqrt{3} \pi \gamma l_{Pl}^2$, $\gamma$ being the Immirzi parameter. The dimensionless volume parameter $v(a)$ is related to the scale factor through the relation
\begin{eqnarray}
v(a) = \left(\frac{6}{8\pi\gamma\l^2_{Pl}} \times (2\pi^2)^{2/3}\right) a^3\,.
\end{eqnarray}
It is sufficient to consider the (LQC) Friedmann equation in this case since the Raychaudhuri equation can be derived from it via a derivative with respect to $\tau$. There are two regions in which the behaviour of this equation is of interest to us. One of them is when we are near the $a\approx 0$ case. Note that one never reaches this limit in the Lorentzian regime, and one concludes there is a bounce considering only Lorentzian histories. However, in order to investigate whether there is a well-defined, non-singular instantonic solution to the Euclidean equations of motion, one needs to examine this limit. For $a \ll 1$, $f(a)\sim a^3$ whereas for $a \gg 1$, $f(a)=1$. Thus, in the (relatively) large volume limit, we get the usual holonomy-modified LQC Friedmann equation, whereas in the other limit, the inverse-triad corrections play a crucial role to make the instanton non-singular.

\begin{figure*}
\begin{center}
\includegraphics[scale=0.75]{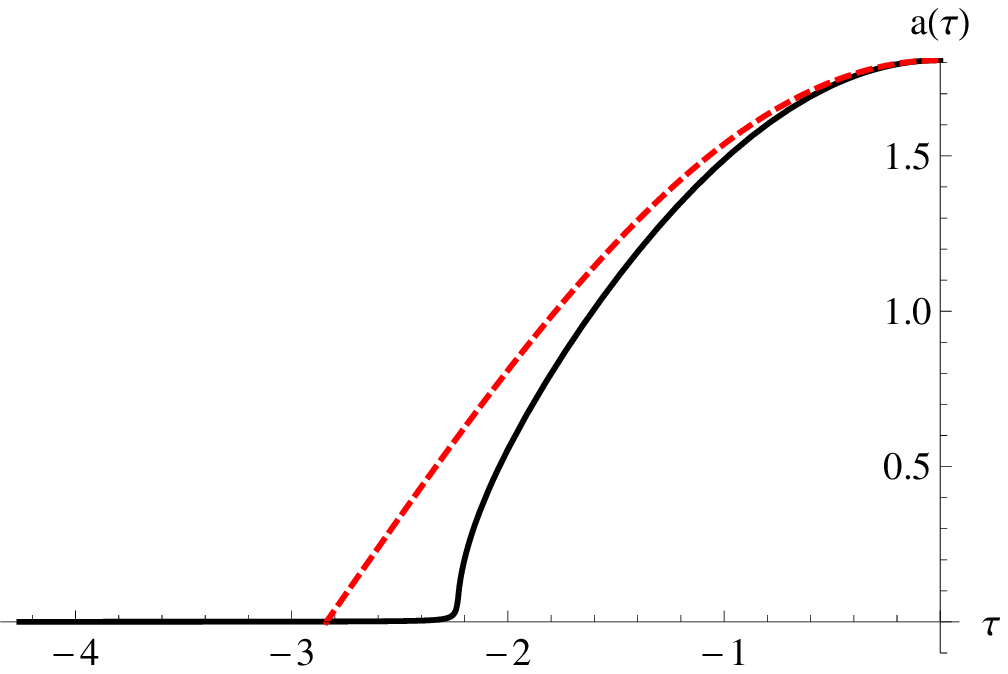}
\includegraphics[scale=0.75]{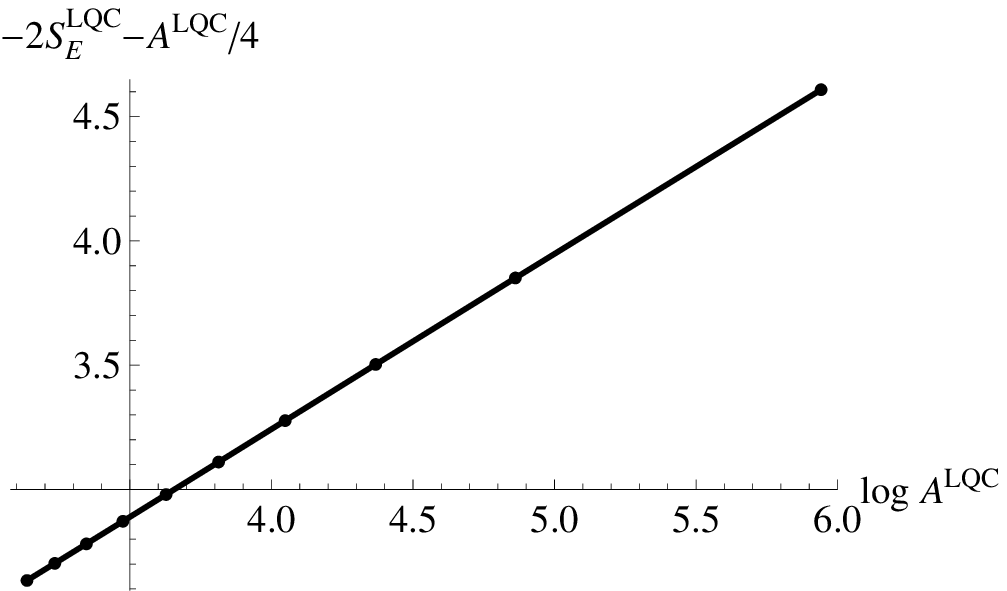}
\caption{\label{fig:dS}Left: a typical solution $a(\tau)$ for $\Lambda = 1$ and $l_{Pl} = 0.1$. The red dashed curve is the Einstein gravity case with the same $\tilde{a}$. Right: $-2S_{\mathrm{E}} - \mathcal{A}/4$ versus $\log \mathcal{A}$. This shows Eqn.~(\ref{eq:ent}).}

\includegraphics[scale=0.75]{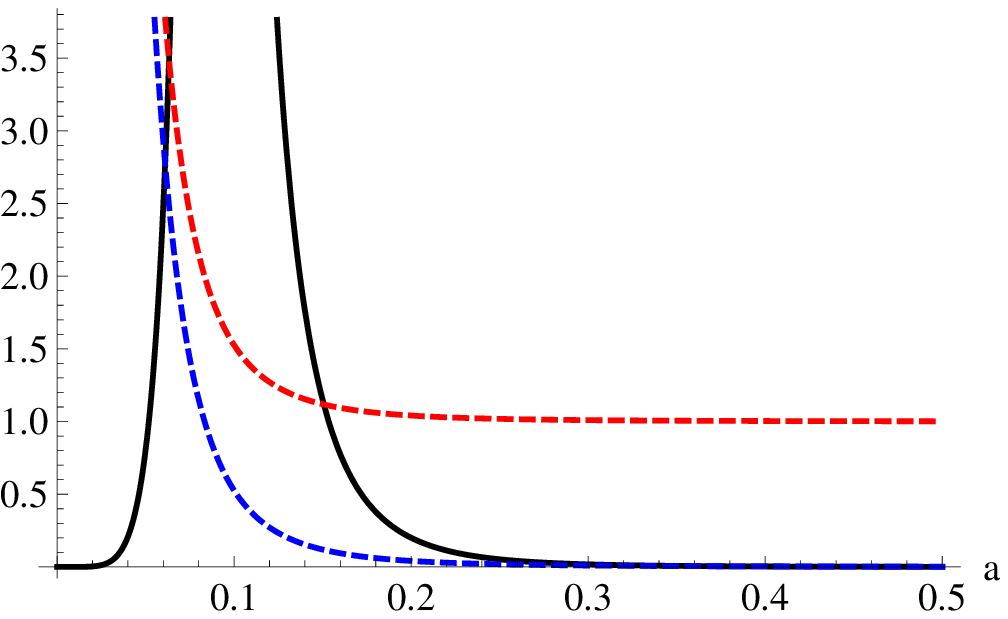}
\includegraphics[scale=0.7]{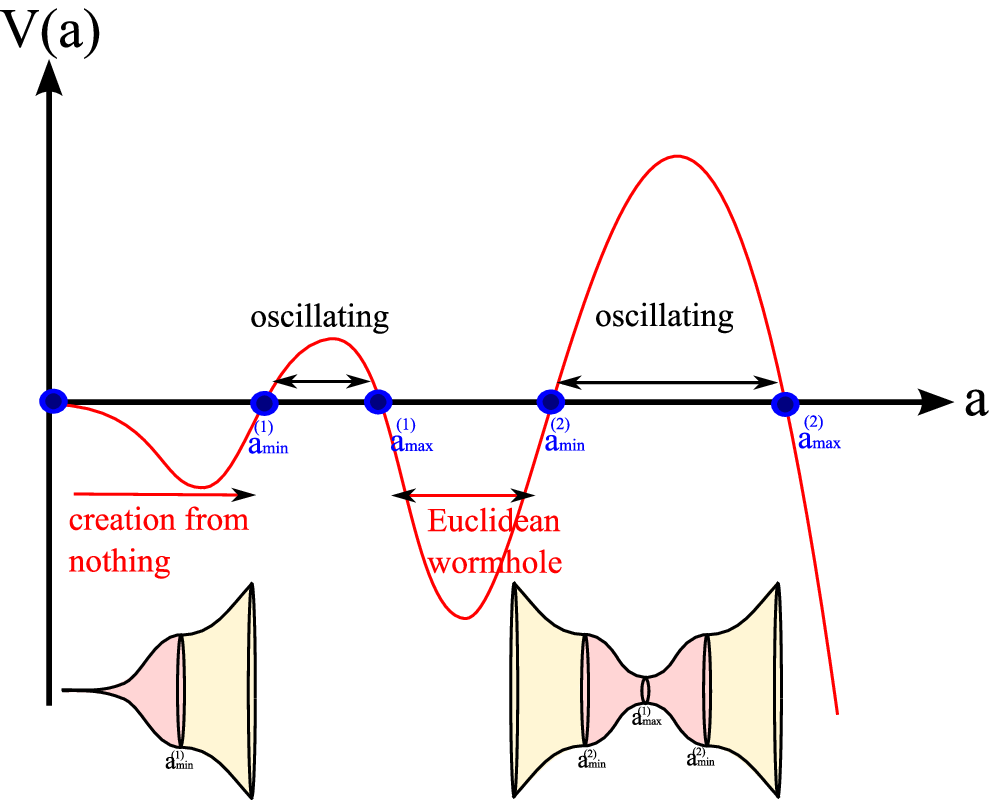}
\caption{\label{fig:sfield}Left: the left hand side of Eqns.~(\ref{eq:zeros1}, \ref{eq:zeros2}) (black) and the right hand side of Eqn.~(\ref{eq:zeros1}) (blue dashed) and Eqn.~(\ref{eq:zeros2}) (red dashed) for $l_{Pl} = 0.1$ and $\dot{\phi}_{0} = 10^{5}$. This shows four zeros of $\mathcal{V}$ in Eqn.~(\ref{eq:sfield}). Right: a conceptual interpretation of $\mathcal{V}$.}
\end{center}
\end{figure*}

The integral defined in Eqns.~(\ref{LQCNB}, \ref{EffectiveAction}), together with Eqn.~(\ref{EffFriedmann}), is taken over a class of paths satisfying the final condition $a(\eta_f) = \tilde{a}$ and the initial condition $a(\tau_i)=0$. For the HH initial state with the EH action, one further gets that $\dot{a}(\tau_i) = 1$ from the Hamiltonian constraint. However, in our case of LQCNB, such an instanton does not exist. If we still want to choose $a(\tau_i) = 0$ (since we wish to have a `regular closed-off geometry'), we get from (\ref{EffFriedmann}) that $\dot{a} = 0$ as well. Thus, we see that not only are the `inverse-triad' corrections, manifested by the functions $f(a)$ \footnote{There are some quantum ambiguities in calculating the exact form of this function but we stick to the choices made in \cite{Ashtekar:2006es} here.}, crucial for a non-singular instanton in this case, but also they lead to a geometrically (quantitatively) different solution from the HH one (left of Fig.~\ref{fig:dS}).

The structure of the effective potential, $\mathcal{V}$, is too complicated due to LQC corrections to evaluate these path integrals analytically, even in the saddle point approximation. Instead, we go on to integrate this numerically, and express the Euclidean (LQC) action as 
\begin{eqnarray}\label{eq:ent}
 	-2 S^\text{LQC}_{\mathrm{E}} \simeq \frac{\mathcal{A}}{4} + c + d \log \mathcal{A} + \dots,
\end{eqnarray}
where $\mathcal{A} = 4 \pi \tilde{a}^{2}$ (right of Fig.~\ref{fig:dS}). Expressing the semiclassical factor in the quantum mechanical amplitude as above, the nucleation probability is
\begin{eqnarray}
	\mathcal{P} \simeq e^{-2 S^\text{LQC}_{\mathrm{E}}}\,.
\end{eqnarray}
The comparison of this probability with the original HH state, with the EH action, reveals additional corrections, with LQC-dependent parameters $c$ and $d$ (more details on numerical calculations in \cite{Upcoming2}). Assuming that the other constants of nature (such as $\gamma$) stay the same, then $c$ can be absorbed away with a change in the normalization whereas the true corrections to the EH result comes from the presence of a positive $d$. This shows that, in LQC, one has a greater probability to have a nucleating dS space. However, similar corrections might also appear from next-to-leading order calculations of the decay rate and, perhaps, the only unambiguous conclusion is that corrections from LQC is small in this case.

\textit{Free scalar field.--} Having shown that there exists a (different) instantonic solution for the well-studied ($k=1$) dS universe for LQC, we now go on to find Euclidean instantons for matter given by a massless scalar field. For the EH action, there exists no such instantonic solutions for the no-boundary wavefunction (in the absence of any scalar-field potential). But the situation is qualitatively different for LQC due to modifications to the equation of motion of the scalar field. Once again, we first write down the effective Friedmann equation for the system as $\dot{a}^{2} = - \mathcal{V}(a)$, with
\begin{widetext}
\begin{eqnarray}\label{eq:sfield}
\mathcal{V} = \frac{8\pi G}{3} a^{2} f^2(a) \left[\frac{a^6 \pi}{4\sqrt{3}\gamma^3 l^6_{Pl}} \left(\frac{\rho}{\rho_c}\right) \left(\frac{g(a)}{f(a)}\right) - \rho_1\right] \left[\frac{1}{\rho_c}\left(\rho_2 - \frac{a^6 \pi}{4\sqrt{3}\gamma^3 l^6_{Pl}}\left(\frac{\rho}{\rho_c}\right) \left(\frac{g(a)}{f(a)}\right)\right) \right],
\end{eqnarray}
\end{widetext}
where
\begin{eqnarray}
g(a) &:=& \frac{27}{8} \huge\big| |v(a)+1|^{1/3} - |v(a) - 1|^{1/3} \big|^3,\\
\rho &:=& \frac{p_\phi^2}{\left(2\pi^2\right)^2 a^6}\,.
\end{eqnarray}
However, in this case, the usual relation between the momenta of the scalar field and its time derivative is also modified as follows
\begin{eqnarray}
p_\phi = \left(\frac{8\pi\gamma l_{Pl}^2}{6}\right)^{3/2} B^{-1}(a) \dot{\phi}
\end{eqnarray}
where
\begin{eqnarray}
B(a) = \frac{2\pi^2}{\sqrt{3}} \left(\frac{8}{27}\right)\left(\frac{6}{8\pi\gamma l_{Pl}}\right)^{3/2} a^3  g(a)\,.
\end{eqnarray}
These inverse-triad corrections ($f(a), g(a), B(a)$) ensure a modification to the scalar-field equation as well
\begin{eqnarray}
\ddot{\phi} - \left(\frac{\dot{B}(a)}{B(a)}\right) \dot{\phi} = 0\,,
\end{eqnarray}
with the general solution given by $\dot{\phi} = \dot{\phi}_{0} B$, with constant $\dot{\phi}_{0}$. Using these two equations, one can derive the Raychaudhuri equation as in GR, albeit the form of it gets modified due to LQG corrections.

Looking at the asymptotic behaviour of the function $g(a)$ defined above, it is straightforward to show that in the large volume limit, one gets that $B\sim 1/a^3$, thereby leading to the usual scalar-field equation
\begin{eqnarray}
\ddot{\phi} +3 H \dot{\phi} = 0\,,	
\end{eqnarray}
where $H:=\dot{a}/a$ is the Hubble parameter. In this case, we get the classical solution, $\dot{\phi}\sim 1/a^3$. On the other hand, when $a\ll 1$, we find that the effects of inverse-triad corrections kick in, setting up an anti-friction term (with an opposite sign), leading to a ``superinflationary'' era. In this case, the equation is of the form
\begin{eqnarray}
\ddot{\phi} - 12 H \dot{\phi} = 0\,,	
\end{eqnarray}
leading to the solution $\dot{\phi} \sim a^{12}$ in this regime. This is crucial for our purposes of getting a well-defined, non-singular instanton in this case as well, thanks to these additional factors of the scale factor coming from the inverse-triad terms. As the scale factor approaches zero, we find the regular instantonic solution at $(a=0, \dot{a} =0)$ like in the dS case. Unlike in the EH theory, we thus get a \textit{new} instantonic solution for the LQC no-boundary wavefunction for a free inflaton field, thereby not requiring an inflationary `slow-roll' type potential term any longer for a smooth beginning in the deep quantum regime.

We have not said anything thus far regarding the initial conditions for the scalar field corresponding to our LQCNB state. For the original HH proposal, in the EH case, one \textit{requires} the condition such that $\dot{\phi}\rightarrow 0$ as $a\rightarrow 0$, for well-behaved instantons (in the presence of some scalar potential). However, in this LQCNB case, we are free to choose any value for $\dot{\phi}_0$. There are, of course, other reasonable restrictions which have to imposed such as having a $p_\phi \gg \hbar$, so that the state remains semiclassical at late times. This condition is important for both the validity of the effective Friedmann equation in LQC as well as our saddle point approximation, and has been enforced in our numerical investigations.

From the numerical solutions, it is possible to observe that the $\mathcal{V}(a)$ has zeros if $a(\tau)$ satisfies one of two equations:
\begin{eqnarray}
\frac{4}{27\sqrt{3}} \frac{\dot{\phi}_{0}^{2}}{\rho_{c}^{2}} \frac{g(a)}{f(a)} &=& -\sin^2 \(\frac{\sqrt{\Delta}}{a}\) + (1+\gamma^2) \frac{\Delta}{a^2},\label{eq:zeros1}\\
\frac{4}{27\sqrt{3}} \frac{\dot{\phi}_{0}^{2}}{\rho_{c}^{2}} \frac{g(a)}{f(a)} &=& \cos^2 \(\frac{\sqrt{\Delta}}{a}\) + (1+\gamma^2) \frac{\Delta}{a^2}.\label{eq:zeros2}\,.
\end{eqnarray}
On examining typical shapes, we see that there are four zeros in general, say $0 < a^{(2)}_{\mathrm{min}} < a^{(2)}_{\mathrm{max}} < a^{(1)}_{\mathrm{min}} < a^{(1)}_{\mathrm{max}}$ (left of Fig.~\ref{fig:sfield}).

The usual interpretation in LQC for such models is that the universe oscillates between $a^{(1)}_{\mathrm{min}}$ and $a^{(1)}_{\mathrm{max}}$, two quantum bounces resolving the big bang and the big crunch singularities \cite{Ashtekar:2006es}. One does not consider oscillations between $a^{(2)}_{\mathrm{min}}$ and $a^{(2)}_{\mathrm{max}}$ since this region is usually assumed to be `forbidden' in LQC, if one considers only Lorentzian histories. However, now considering Euclidean trajectories, we get two possible scenarios (right of Fig.~\ref{fig:sfield}):
\begin{enumerate}
	\item The universe with size $a^{(2)}_{\mathrm{min}}$ is created from nothing. It begins oscillations between $a^{(2)}_{\mathrm{min}}$ and $a^{(2)}_{\mathrm{max}}$. At a certain time, the universe tunnels from $a^{(2)}_{\mathrm{max}}$ to $a^{(1)}_{\mathrm{min}}$. Eventually, it begins the second oscillation between $a^{(1)}_{\mathrm{min}}$ and $a^{(1)}_{\mathrm{max}}$, as is usually expected in LQC.
	\item Two universes with size $a^{(1)}_{\mathrm{min}}$ are created from nothing via an Euclidean wormhole \cite{Chen:2016ask,Kang:2017jmq} which starts from $a^{(1)}_{\mathrm{min}}$, decreases to $a^{(2)}_{\mathrm{max}}$, and bounces back to $a^{(1)}_{\mathrm{min}}$ via (real) Euclidean time.
\end{enumerate}

\textit{Conclusion.--} One finds that LQC has a small contribution in the dS case while a substantially new result for the free scalar field. This is because in the former case, already in EH theory, there exists a potential barrier with a well-defined Euclidean instanton. Thus, LQC simply modifies the limits of this potential barrier for the universe to nucleate out of and results in a small correction to the EH result. In the free scalar field case, there exists no solutions to the (complex) equations of motion within the no-boundary proposal, for the EH case. However, for the LQCNB state, there exists a well-defined (real) Euclidean instanton even in the massless inflaton case, which corresponds to finite probabilities for a universe nucleating from nothing, or via a Euclidean wormhole.

The original no-boundary wavefunction was an attempt to formulate initial conditions for inflation as a topological principle; therefore, being applicable to a variety of fundamental quantum gravity approaches. In this article, we show that such a `natural' initial condition \cite{Hartle:1983ai}, complemented by nonlocal (on the Planck scale) quantum geometry regularizations, result in a non-singular quantum gravity extension for inflation which is far more general than the original proposal. We find that there is an expanded range of initial conditions for the (momenta of the) inflaton field, allowing for a much larger class of potentials for setting up inflation. We study two extreme cases -- pure dS and a massless inflaton field, with the latter being an example of new solutions for the no-boundary wavefunction in LQC, having well-defined Euclidean saddle-points where the EH theory did not. This opens up possibilities for a wide range of initial values, within the no-boundary proposal, leading to inflation due to necessary (loop) quantum gravity corrections to the EH action in deep quantum regimes. In other words, LQC corrections to the no-boundary proposal result in removing restrictions on the shape of the potential required for starting inflation, such as having a false vacuum.

On the LQC side, these type of proposals give us a new principle for choosing an initial state as well as a different way to understand singularity-resolution. For instance, the universe could have been created from nothing in LQC, in addition to the usual bounce paradigm, with a natural interpretation of this as a superposition of histories with different probabilities. Moreover, the no-boundary state uncovers new physical phenomenon in the Planckian era, such as the (Lorentzian) region between $a^{(2)}_{\mathrm{min}}$ and $a^{(2)}_{\mathrm{max}}$ in the free scalar field case. In this case, one can even consider scenarios with two arrows of time leading to rich possibilities for quantum cosmology. As we have shown, such a choice for smooth initial conditions is consistent with the conclusions for LQC background evolution at times soon after the `creation' of the universe, both in the dS and the free scalar field case. Given this principle for choosing the initial state, one can perhaps distinguish between the different interpretations mentioned above by going beyond homogeneous minisuperspace backgrounds, which we set aside for future investigation.

\smallskip

\noindent{\bf Acknowledgements:}
This work was supported in part by the Korean Ministry of Education, Science and Technology,
Gyeongsangbuk-do and Pohang City for Independent Junior Research Groups at
the Asia Pacific Center for Theoretical Physics.

\bibliographystyle{apsrev4-1}
\bibliography{refs}

\end{document}